# High-energy mid-infrared sub-cycle pulse synthesis from a parametric amplifier

**Houkun Liang[1,2], Peter Krogen[1], Zhou Wang[3], Hyunwook Park[3], Tobias Kroh[1,4], Kevin Zawilski[5], Peter Schunemann[5], Jeffrey Moses[1,6], Louis F. DiMauro[3], Franz X. Kärtner[1,4,7], and Kyung-Han Hong[1,*]**

[1]*Department of Electrical Engineering and Computer Science and Research Laboratory of Electronics, Massachusetts Institute of Technology (MIT), Cambridge, Massachusetts 02139, USA*

[2]*Singapore Institute of Manufacturing Technology, 2 Fusionopolis Way, Singapore 138634*

[3]*Department of Physics, The Ohio State University, Columbus, Ohio 43210, USA*

[4]*Center for Free-Electron Laser Science, DESY and Department of Physics, University of Hamburg, D-22607 Hamburg, Germany*

[5]*BAE System, MER15-1813, P.O. Box 868, Nashua, New Hampshire 03061, USA*

[6]*School of Applied and Engineering Physics, Cornell University, Ithaca, New York 14853, USA*

[7]*The Hamburg Center for Ultrafast Imaging, Luruper Chaussee 149, 22761 Hamburg, Germany*

**High-energy, carrier-envelope phase (CEP)-stable, sub-cycle, mid-infrared (mid-IR) pulses can provide unique opportunities of exploring phase-sensitive strong-field light-matter interactions in atoms, molecules, and solids. In the mid-IR wavelength, the ponderomotive energy of laser pulses is dramatically increased (versus the visible/near-infrared) and, therefore, the Keldysh parameter is much smaller than unity even at relatively modest laser intensities. This enables to study the sub-cycle electron dynamics in solids via high-harmonic generation (HHG)[1-4] without damage. One can also control the electron emissions from nano-devices in the sub-cycle time scale[5-7]. These efforts are opening a great opportunity towards "petahertz electronics"[8,9]. Here, we present a high-energy, sub-cycle pulse synthesizer based on a mid-IR optical parametric amplifier (OPA), pumped by CEP-stable, 2.1 μm femtosecond pulses, and its application to HHG in solids. The signal and idler combined spectrum spans from 2.5 to 9.0 μm, which covers the whole midwave-infrared (MWIR) region. We coherently synthesize the passively CEP-stable few-cycle signal and idler pulses to generate 33 μJ, 0.88-cycle**

**(12.4 fs), multi-GW pulses centered at ~4.2 μm, which is further energy scalable. The in-line synthesis of the CEP-stable sub-cycle pulse is realized through the type-I collinear OPA with minimal temporal walk-off. The MWIR sub-cycle pulse is used for driving HHG in thin silicon samples, producing harmonics up to ~19th order with a continuous spectral coverage due to the isolated emission by the sub-cycle driver. Our demonstration offers an energy scalable and technically simple platform of laser sources generating CEP-stable sub-cycle pulses in the whole MWIR region for investigating isolated phase-sensitive strong-field interactions in solids and gases[10-13].**

Generation of high-energy, few-cycle mid-IR pulses has progressed dramatically over the last decade, driven by a number of applications, such as coherent soft X-ray HHG[14-16], incoherent hard X-ray generation in laser-induced plasmas[17], sub-femtosecond electron emission[6], two-dimensional infrared spectroscopy[18], and time-resolved imaging of molecular structures[19]. The use of a CEP-stable single-cycle or even sub-cycle pulse can inherently isolate the electron dynamics in the strong-field interactions[12,13]. Therefore, high-energy, CEP-stable, sub-cycle mid-IR pulses can be a very unique tool for investigating ultrafast dynamics of strong-field interactions in solids and gases. Some examples are the sub-cycle control of electron motions via HHG in solids[2-4], the sub-cycle electron tunneling in nano-devices[7], the generation of isolated attosecond[10] or even zeptosecond X-ray pulses[16], controlling strong-field molecular ionization and dissociation[20], steering the atomic-scale motion of electrons[12], and sub-femtosecond control and metrology of bound-electron dynamics in atoms[13]. Recently, intensive effort has been made to generate few-cycle mid-IR pulses using several techniques, such as OPA [21, 22], laser filamentation[23] and difference-frequency generation (DFG)[24, 25]. Chirped quasi-phase matching gratings were employed in some cases to achieve broad phase-matching bandwidth[22, 25]. Self-compression after spectral broadening in dielectric materials[26-29] has also been used for generating sub-two-cycle mid-IR pulses. Regarding the generation of mid-IR sub-cycle pulses, four-wave mixing through filamentation in a gas[30] and a technique that cascades DFG, spectral broadening, and chirp-compensation[31] have been demonstrated. However, the former gives low pulse energy (~0.5 μJ)[32] and limited energy scalability as well as conical emission, while the latter has unknown CEP stability due to the complex nonlinear processes involved besides the low energy (~1 μJ). These impose limitations in the applications to strong-field light-matter interactions. On the other hand, in the visible to short-wavelength IR range, high-energy sub-cycle pulses have been demonstrated using pulse synthesizers[12, 13, 33, 34]. Greater-than-octave-spanning spectra are generated and amplified in different spectral bands. After the phase management of individual bands, the multi-color pulses are coherently combined to form a sub-cycle pulse. The main challenge of this approach is the complexity of the system because sophisticated phase control and stabilization have to be implemented for eliminating the relative timing and phase jitters from the individual bands.

Coherent synthesis between the signal and idler of a type-I collinear OPA in a passive way is an intriguing alternative for sub-cycle pulse generation. With near-degenerate signal and idler wavelengths, the total OPA bandwidth can be huge in a collinear geometry when the group velocity dispersion at the degenerate wavelength is small[35]. The signal and idler pulses are tightly synchronized in an OPA by nature. However, this method is challenging in the visible/near-IR range for several reasons. First, the dispersion over multi-octave bandwidths is very large, such that the signal and idler pulses will require post-compression. Second, the signal and idler need to have a stable relative CEP, which can only be achieved when both the pump and signal pulses are CEP stable, as their phase difference is transferred to the idler CEP. In the visible and near-IR ranges, such schemes usually require active CEP stabilization. In contrast, in the mid-IR, the dispersion of OPA media can be very low within the transmission window, which enables the direct synthesis of the signal and idler pulses without post compression. Moreover, in the scheme described below, the 2.1-μm pump laser itself is a passively CEP-stabilized, optical parametric chirped-pulse amplifier (OPCPA), and its pulse width is short enough to pump a white-light generation (WLG) stage and generate an octave-spanning CEP-stable signal pulse for a mid-IR OPA.

Here, we present a multi-gigawatt sub-cycle mid-IR pulse synthesizer based on in-line multiplexing of the signal and idler pulses from an OPA. Furthermore, we demonstrate HHG in silicon samples up to ~19[th] order and observed a continuous spectrum due to the sub-cycle driver pulses. The signal and idler pulses covering 2.5 to 4.4 μm and 4.4 to 9.0 μm, respectively, are synthesized automatically in the collinear type-I OPA, due to the minimal dispersion and temporal walk-off of a thin $CdSiP_2$ (CSP) crystal, which also supports a phase-matching bandwidth greater than one octave at the idler wavelength (see Fig. S1 for the phase-matching curve in the Supplementary Materials). The CEP-stable pump from the 2.1-μm OPCPA at 1 kHz repetition rate provides the CEP-stable signal for the mid-IR OPA, via WLG. The CEP of the idler pulse is also passively stabilized by DFG-like parametric process between the WLG signal and the pump pulses, regardless of the CEP stability of the pump. Therefore, the CEP stability of both signal and idler pulses is ensured without active stabilization, as confirmed by the *f*-3*f* spectral interferometry measurements over more than 6 minutes. Synthesized pulses with 33 μJ energy and multi-GW peak power having 12.4 fs full-width at half maximum (FWHM) duration, characterized using a cross-correlation frequency-resolved optical gating (XFROG) device, are demonstrated.

The scaling of energy and peak power is relatively straightforward by adding more OPA stages with higher pump energy. The availability of Joule-level ps Yb-doped lasers[36] ensures the energy scalability of our 2.1 μm OPCPA to multi-ten mJ level, which can eventually increase the synthesized mid-IR pulse energy to multi-mJ. It is worth mentioning that the mid-IR crystals like CSP and ZnGeP$_2$ (ZGP) have damage threshold greater than 200 GW/cm$^2$ pumped by femtosecond ~2-μm pulses. The availability of >20 mm × 20 mm crystal aperture[37] allows the employment of a ~2-μm, ~40-fs pump with >20-mJ pulse energy to achieve the mJ sub-cycle mid-IR pulse amplification, as discussed in Fig. S2 in the Supplementary Materials. Recently, there have been mid-IR OPCPA demonstrations with sub-mJ energy and 4–8 cycle pulse width, pumped by high-energy ~2 μm picosecond Ho-doped lasers.[38, 39] Our approach can be combined with the advanced ~2-3 μm pump laser technologies[38-40] to realize a relatively compact multi-mJ near-single-cycle mid-IR laser source if the sub-50 fs duration can be accessed by these lasers via external pulse compression techniques[29]. On the other hand, our passive synthesis scheme can be widely adopted for generating sub-1.5-cycle, sub-mJ mid-IR pulses if existing ~2 μm OPA systems pumped by conventional Ti:sapphire laser amplifiers are used for pumping CSP and ZGP crystals.

**Results**

**Experimental setup of mid-IR OPA.** The schematic of the mid-IR sub-cycle pulse generation and characterization is presented in Fig. 1. An octave-spanning Ti:sapphire oscillator provides the seed for the 2.1-μm OPCPA through an intrapulse DFG stage which ensures passive CEP stabilization. The kHz, multi-mJ, CEP-stable, 2.1-μm OPCPA[41] serves as the pump of the mid-IR OPA. A 20 μJ portion of the 2.1-μm pump is split for WLG in a 6-mm-thick BaF$_2$ plate[27] as the signal of the mid-IR OPA. A 1.1-mm-thick CSP is chosen for the type-I parametric conversion for its large nonlinear coefficient, broad phase-matching bandwidth, and high damage threshold pumped by 2.1-μm pulses. The pump beam size is ~6 mm in diameter and the intensity is 150 GW/cm$^2$ at 800 μJ of pump energy. The output pulse from the mid-IR OPA together with a ~10 μJ, 2.1-μm pulse split from the pump which serves as the reference beam are sent into the XFROG for the temporal characterization. More details of the setup are found in Methods section.

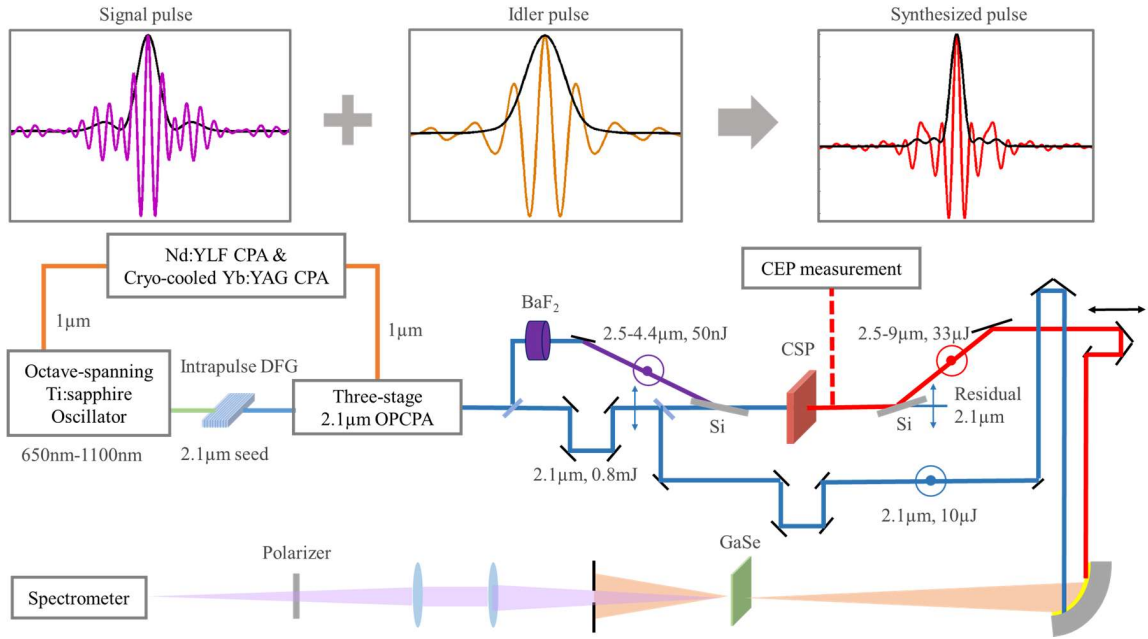

**Fig. 1.** The schematic of the high-energy phase-stable sub-cycle mid-IR OPA. CPA: chirped-pulse amplifier. Polarizations of the beams are marked by double-headed arrows and concentric circles. 300-μm thick Si wafers at Brewster angle are used as polarization beam splitter and beam combiner to transmit the 2.1-μm pump pulse and reflect the signal and idler pulses. The synthesized pulses and a branch of 2.1-μm reference pulses are sent into XFROG with 30-μm thick GaSe nonlinear crystal. The synthesis of a sub-cycle mid-IR pulse from coherently combining the sub-2-cycle signal and idler pulses is shown conceptually on the top of the figure.

Fig. 2(a) shows the mid-IR spectrum of the signal pulse which is a part of the supercontinuum from the WLG. We obtain ~50 nJ of pulse energy within the spectral window of 2.5– 4.4 μm. The spectra of the amplified signal and idler with 0.8 mJ pump energy are shown in Fig. 2(b), spanning from 2.5 to 9.0 μm. The dip at ~3 μm originates from the WLG seed. It is worth noting that the long-wavelength component of the pump at 2.2–2.3 μm gives excellent phase matching in the spectral range of 4–8 μm, as shown in Fig. S1(b) in the Supplementary Materials. Therefore, the 2.2–2.3 μm wavelength component of our broadband pump is very helpful to achieve the octave-spanning parametric conversion. 33 μJ output energy from the mid-IR OPA at 0.8 mJ pump is demonstrated as shown in Fig. 2(c), of which there is a 12 μJ idler pulse spanning from 4.4 to 9.0 μm. The conversion efficiency to the synthesized pulse is ~6% considering the Fresnel reflection of the pump at the uncoated CSP crystal. While the available pump energy is higher than 2 mJ, the OPA stage has been designed at ~1 mJ of pump energy in this work because of the energy loss from metallic mirrors (13 bounces). The shot-to-shot energy stability of the mid-IR OPA is ~2.7% rms (10,000 shots) with an oscillation of ~7% peak-to-peak over every ~10 minutes of period, which is attributed to the on/off operation of the air conditioner in the lab, as shown in Fig. S9 of the Supplementary Materials. There is no noticeable degradation of energy over several hours. The near-Gaussian signal and idler beam profiles, measured with a pyroelectric camera (PyroCam III, Spiricon) are presented in Fig. 2(d) and (e), respectively. It should be noted that the idler spectrum can be even broader (up to 10 μm along with a stronger signal pulse), if we use a higher 2.1 μm pulse energy for WLG, as shown in Fig. S3 in the Supplementary Materials. However, we limit the energy to ~50 nJ within the signal bandwidth because the CEP of the signal generated by WLG is found to be less stable at higher energy.

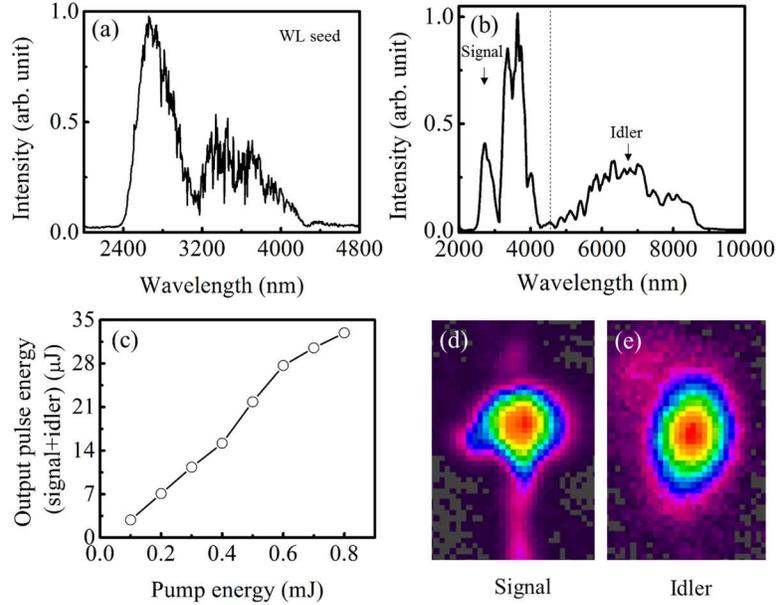

**Fig. 2** (a) The signal spectrum from WLG at $BaF_2$, measured after a 2400 nm long-pass filter. (b) The measured output spectrum of the mid-IR OPA. The dotted line separates the signal and idler spectra. (c) The output pulse energy of the mid-IR OPA vs. pump energy. The far-field beam profile of the amplified signal and idler is presented in (d) and (e), respectively.

**Characterization of CEP-stable few-cycle signal and idler pulses.** The temporal profile of the 2.1-μm pulses from the OPCPA, which serves both as the pump of the mid-IR OPA and as reference of the XFROG, is characterized using a second-harmonic generation (SHG) FROG apparatus. As shown in Figs. 3(a)-(d), the pump pulse whose spectrum spans from 1.8 μm to 2.3 μm is within 5% of its transform limit, with a pulse width of 26 fs in FWHM. The signal pulse from WLG has a slight self-compression in $BaF_2$ which has small anomalous dispersion at 2.1 μm[27]. The pulse duration of the amplified signal is measured to be ~20 fs using the SHG FROG, as shown in Fig. 3(e).

Before we explore the coherent pulse synthesis, we investigate the temporal profile of the idler pulse which already contains an octave-spanning spectral content. The amplified idler pulse is independently characterized using a home-built mid-IR second-order interferometric autocorrelator. An uncoated 1-mm-thick ZnSe plate is employed as a beam splitter, and a 4.5-µm long-pass filter is used to isolate the idler pulse from the signal and any residual pump. The anomalous dispersion from the ZnSe plate is well compensated by the normal dispersion from the Ge substrate of the long-pass filter, however there is approximately 8000 fs$^3$ of uncompensated third-order dispersion (TOD) in this configuration. As shown in Fig. 3(f), the measured interferometric autocorrelation trace is plotted together with a transform-limited trace calculated using the measured spectrum. The good agreement between the 3 central lobes of the pulses manifests that the idler is nearly transform limited with a pulse width less than 1.5 optical cycles, centered at 6.4 µm. ~31 fs pulse width is deconvoluted assuming the idler pulse is in the Gaussian profile. The unsuppressed pedestal from the measured pulse accounts for the uncompensated TOD. This measurement shows that the octave-spanning, ~6.4-µm, sub-1.5-cycle idler pulse, which is CEP stable as will be discussed again, is already usable for phase-sensitive strong-field experiments in solids and nano-structures as a stand-alone mid-IR source.

In the collinear type-I OPA, the signal and idler pulses have the same polarization and can be synthesized automatically if the temporal walk-off between them is much smaller than an optical cycle and the CEP of both pulses is stable. The temporal walk-off within the 1.1-mm-thick CSP crystal is as small as ~5 fs (less than a quarter cycle of the idler wavelength, ~6.4 µm). The electric field of the synthesized pulse is calculated as shown in Fig. 3(g), using the measured spectra and the calculated 5 fs temporal walk-off. Here we assume both signal and idler pulses are transform limited, and the CEP of both pulses is zero. 11 fs in FWHM of the synthesized pulse is then obtained as shown in Fig. 3(h), corresponding to 0.8 optical cycles centered at 4.1 µm. This clearly reveals the feasibility of sub-cycle pulse synthesis. An independent simulation also supports this calculation, as shown in Fig. S2 of the Supplementary Materials, along with the energy scaling to mJ level from the second stage OPA. It should be noted that in the case of multi-stage OPA the pulse synthesis always occurs in the final stage. The temporal walk-off in the first or intermediate stage is ignored because we use the idler only from the final stage for the pulse synthesis.

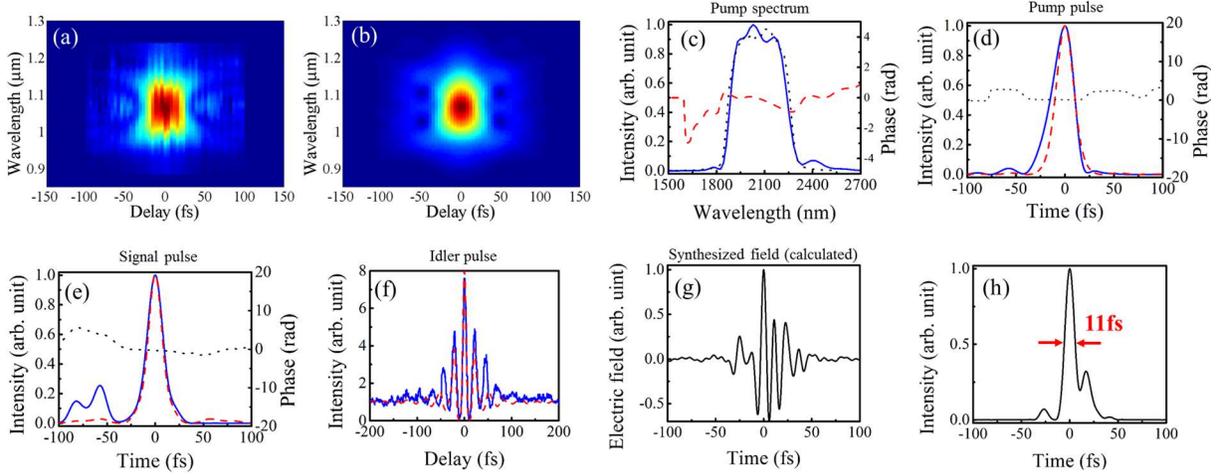

**Fig. 3** (a)-(d) The temporal characterization of the 2.1-μm pump pulse using the SHG FROG. The measured (a) and retrieved (b) FROG traces. (c) The retrieved spectrum (blue solid) and phase (red dash) in the FROG measurement. The separately measured spectrum is shown in black dot which agrees well with the retrieved spectrum in FROG. (d) The retrieved (blue solid) and the calculated transform-limited (red dash) intensity profiles. The black dotted curve is the retrieved phase. The pulse duration of 26 fs FWHM of the pump is measured. (e), (f) The temporal characterization of the amplified signal and idler pulses using the SHG FROG and interferometric autocorrelator, respectively. (e) The retrieved (blue solid) and the calculated transform-limited (red dash) intensity profiles of the amplified signal pulses. The black dotted curve is the retrieved phase. A pulse duration of 20 fs in FWHM of the amplified signal is obtained. See Fig. S7 in the Supplementary Materials for the measured and retrieved SHG FROG traces. (f) The measured autocorrelation trace of the idler pulse (blue solid) and the calculated autocorrelation trace from the measured spectrum assuming the pulse is transform limited (red dash). A pulse duration of 31 fs FWHM of the idler is obtained, assuming Gaussian temporal profile. (g) The calculated electric field of the synthesized mid-IR pulse using the measured spectra of the signal and idler pulses. It is assumed that both the signal and idler pulses are transform limited, the idler leads the signal by 5 fs, and the CEP is zero for both pulses. (h) shows the calculated intensity profile. The calculated duration of the synthesized pulse is 11 fs in FWHM corresponding to 0.8 cycle at 4.1 μm.

The stable CEP of individual pulses as well as the relative phase is crucial for the coherent pulse synthesis in the single-cycle limit. The CEP stability of the 2.1-μm pump is measured using the self-referencing *f-3f* spectral interferometry (SI) of the white light which also serves as the signal of the mid-IR OPA. In the other word, this *f-3f* measurement directly provides the CEP stability of the signal pulses. The third harmonic (TH) of the ~2 μm portion of the signal and the ~680 nm portion of the white light are spectrally interfered, as previously demonstrated[34, 42]. Fig. 4(a) shows the stable *f-3f* fringes over 10 minutes with a 3-shot average, and the shot-to-shot CEP jitter of the signal is measured as ~220 mrad rms. Similarly, the CEP stability of the idler pulse is measured using the cross-referencing *f-3f* SI, which is obtained by interfering the polarization-rotated ~2.1-μm pump and the TH of the idler at ~6.4 μm. The details are shown in the Method section and in Fig. S5 of the Supplementary Materials. The fairly stable single-shot interference fringes were observed over 6 minutes, as shown in Fig. 4(b), owing to the passively CEP-stable nature of the white-light-seeded OPA. These two SI measurements indicate that the pump, signal, and idler all have shot-to-shot stable CEP. The measured phase jitter in Fig. 4(b) is the square root of the sum of the squares of the pump and the idler CEP jitters, i.e. $\sigma_{Measured} = \sqrt{\sigma_{Pump}^2 + \sigma_{Idler}^2}$ because they are not correlated. The shot-to-shot CEP jitter of the idler pulse is then calculated as 270 mrad rms over 6 minutes. It is worth mentioning that the absolute CEP value of the pump pulse and the synthesized pulse can be individually controlled using thin wedge pairs. For example, we can use a ~50 μm-thick wedge pair made of Caesium iodide which has very flat dispersion from 2 to 10 μm. While the passively stabilized CEP of both signal and idler pulses is expected to be maintained beyond ~10 minutes of our measurement, the active compensation of potential CEP drift due to the thermal fluctuations in the lab, as evidenced by the energy measurement in Fig. S9 of the Supplementary Materials, using thin wedges could further improve the long-term CEP stability. The CEP of the synthesized waveform could be directly measured using a recently demonstrated electro-optic sampling method[43].

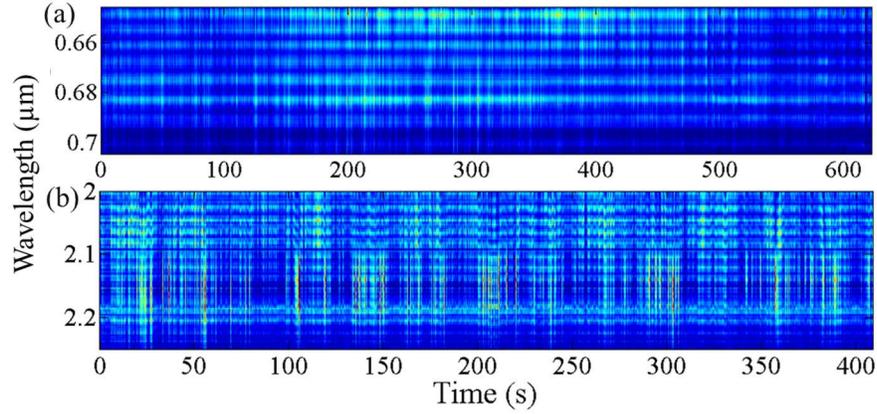

**Fig. 4** (a) The measured self-referencing *f-3f* SI of the 2.1-μm pump pulse over 10 minutes (3-shot average). 220 mrad shot-to-shot CEP jitter is measured over 10 minutes. (b) The measured cross-referencing *f-3f* SI of the idler pulse, between the polarization-rotated ~2.1-μm pump and the TH of the idler at ~6.4 μm over 6 minutes. 270 mrad shot-to-shot CEP jitter of the idler pulse over 6 minutes is calculated with the measured phase jitters of pump and the cross-referencing SI.

**Sub-cycle pulse generation and characterization.** Finally, the temporal profile of the synthesized pulse is characterized with XFROG. The time delay between the signal and pump pulses is finely tuned using a piezo stage within a total delay of ~30 fs, such that the shortest duration is obtained while the amplified energy is maintained at maximum. The measured and retrieved XFROG traces are shown in Fig. 5(a) and (b), respectively, with 1.8% FROG error. The retrieved spectrum shown in Fig. 5(c) agrees well with the measured spectrum in Fig. 2(b). There is minor discrepancy with the spectral edge at 2.5-3 μm of the retrieved spectrum due to the phase matching edge of the GaSe crystal in the sum-frequency generation, as shown in Fig. S6 in the Supplementary Materials. The synthesized pulse has a near-transform-limited main peak and rippling wings as shown in Fig. 5(d), attributed to the interference of the signal and idler pulses. The synthesized pulse duration is measured as ~12.4 fs in FWHM centered at ~4.2 μm, corresponding to 0.88 optical cycle, which is within 10% of its transform limit as shown in Fig. 3(h). With the energy portion into the main pulse of ~70% the peak power reaches ~1.9 GW. We obtain a similarly broad spectrum and high energy from an OPA with a 0.5-mm-thick ZGP crystal, as shown in Fig. S4 in the Supplementary Materials, which is also similar to an early demonstration of ultrabroadband mid-IR OPA with a 12-mm-thick ZGP crystal[44]. However, due to a relatively large temporal walk-off of ~11 fs between the signal and idler with our ZGP crystal, it is less favorable than the CSP crystal for the sub-cycle pulse synthesis.

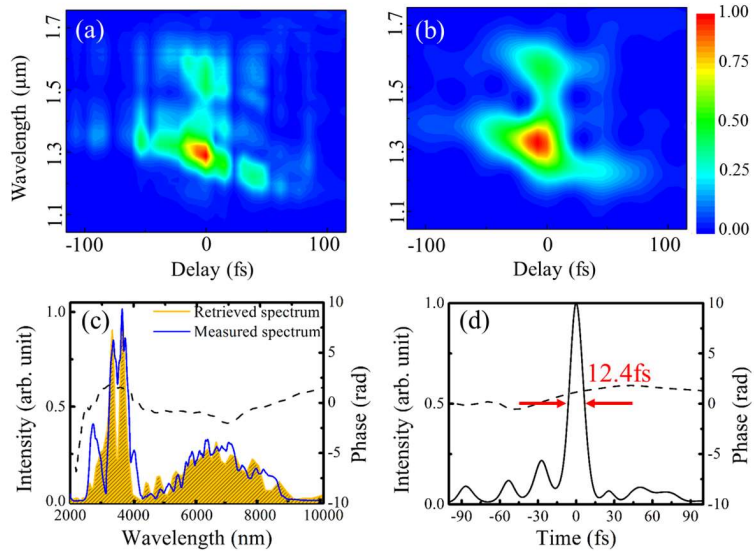

**Fig. 5** The temporal profile characterization of the synthesized mid-IR pulse using XFROG. The measured (a) and the retrieved (b) XFROG traces. The FROG error is 1.8%. The retrieved spectral (c) and temporal (d) intensity profiles of the synthesized pulse. The dotted curves are the retrieved phase. 12.4 fs in FWHM pulse width is measured with a center wavelength at 4.2 μm. It corresponds to 0.88 optical cycle.

**HHG in solids.** We have used the synthesized mid-IR laser pulses to drive HHG in silicon (Si) to show the potential of this source for sub-cycle electron control in solids. A free-standing 200-nm-thick Si (<100>) sample and a 500-nm-thick Si (<100>) sample on a 0.5-mm-thick sapphire substrate are used for generating high harmonics. The OPA signal and idler beams with a Gaussian beam diameter of ~5.5 mm are focused using an $f$=25.4 mm gold-coated off-axis parabolic mirror. The HHG signal is collected using an ultraviolet (UV)-enhanced aluminum-coated off-axis parabolic mirror and spectrally resolved using a visible-to-UV monochromator with an intensified charge-coupled device (ICCD). It should be noted that due to the large difference in the center wavelength between the signal (~3.2 μm) and the idler (~6.4 μm) beams, the focused beam size of the two beams with the same focal length differs by a factor of ~2. Along with the 2 times higher energy of the signal beam than the idler beam, the HHG is dominated by the signal pulse at the focus (z=0 mm). To observe the clear contribution of the idler pulse, we have acquired most high-harmonic spectra at z=0.5–1.0 mm after the focus where the spot sizes of the signal and idler beams are comparable. The beam size in $1/e^2$ radius at z=0.5 mm is estimated as ~57 μm and ~64 μm for the signal and idler, respectively. The estimated intensity of the ~20 μJ synthesized pulse with 60 μm of beam waist is ~$9 \times 10^{12}$ W/cm$^2$, corresponding to the electric field strength of 0.8 V/Å. More details on the HHG setup are found in the Method section.

In the experiment we observed a continuum-like harmonic spectrum using synthesized sub-cycle drive pulses, which will be discussed at the end. To verify that the spectrum originates from HHG in Si, we have generated harmonics with a clear comb structure using few-cycle pulses and demonstrated 4-fold symmetry of the harmonic yields, which reflects the crystal symmetry of Si. This is a feature of HHG rather than incoherent light emission such as fluorescence. A ~16 μJ, few-cycle pulse is obtained by positively chirping the synthesized pulse using a 0.5-mm-thick Si filter (IPA3000, EOC, Inc.) with ~80% transmission over 3–11 μm. The positive dispersion of the Si filter moderately broadens the pulse duration to ~43 fs in FWHM of the double peaks, which is long enough to generate discrete harmonics. The measured OPA spectrum with the filter and the calculated temporal profile are shown in Fig. S9 of the Supplementary Materials. The intensity of the chirped pulse is estimated to be ~$2 \times 10^{12}$ W/cm$^2$ corresponding to an electric field strength of 0.4 V/Å. Figures 6(a) and (b) show the harmonic spectra generated using the ~43 fs chirped pulses in the 200-nm-thick and 500-nm-thick Si samples, respectively. Both odd and even harmonics of ~4.6 μm (black solid lines) are observed up to ~19$^{th}$ harmonic (~244 nm) because the two-color driving field unbalances the electron trajectories and breaks the symmetry[4]. For comparison, the odd harmonics of ~3.2 μm from signal-only driven HHG is also plotted as the blue dotted line of Figs. 6(a) and (b), which are obtained at the beam focus (z=0 mm) where the signal pulse dominates HHG over the idler pulse due to the smaller beam size (half) and higher pulse energy (twice). It is noted that the spectral comb structure is more clearly observed in Fig. 6(a), which is potentially due to the effect from the sample thickness or the substrate, and requires more investigation. The spectral intensity as a function of the crystal orientation, driven by the ~43 fs chirped pulse with the 500-nm-thick Si sample, in Fig. 6(c) clearly shows the 4-fold symmetry of Si when the sample is rotated around the <100> crystal axis, confirming that the emission spectrum originates from HHG in Si. Finally, Fig. 6(d) shows the harmonic spectrum driven by the synthesized sub-cycle pulse without the 3–11 μm Si filter. The harmonic comb structure is almost washed out and a near-continuous spectrum is observed due to the sub-cycle nature of the driver pulse that isolates the harmonic emission. Further investigations on the CEP dependence along with the pump-probe analysis will enable the observation and control of isolated sub-cycle electron dynamics.

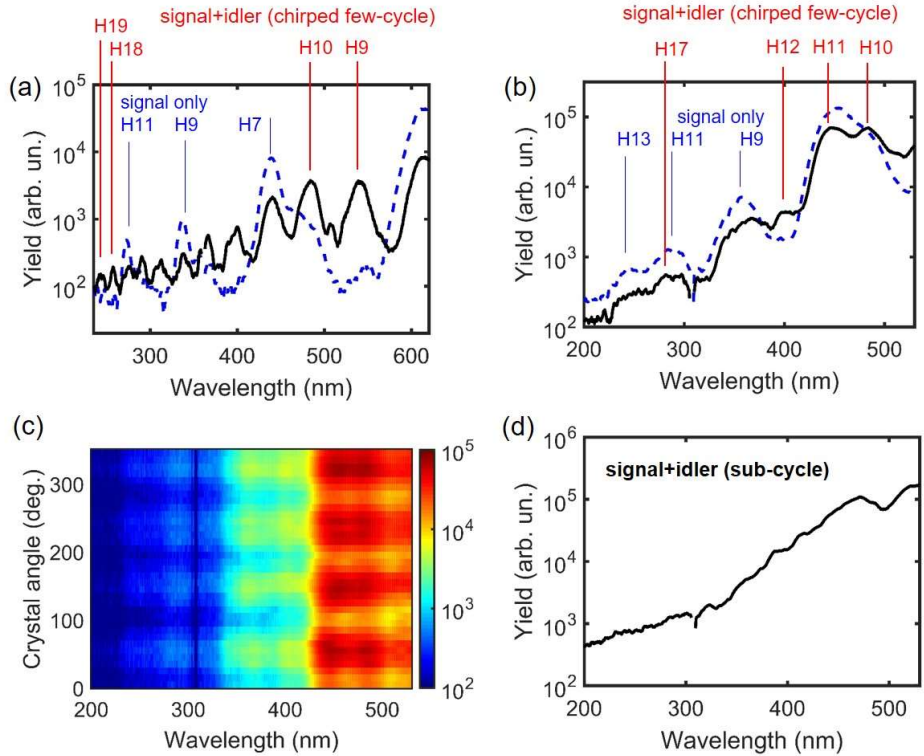

**Fig. 6** Mid-IR-driven HHG in silicon samples. (a) Harmonic spectra generated in a 200-nm-thick free-standing Si sample. The solid black line represents the spectrum generated by a chirped, ~43 fs, synthesized pulse, where the odd and even harmonics up to 19$^{th}$ order are observed. The blue dotted line represents the spectrum generated by only the signal pulse, where only odd harmonics up to 11$^{th}$ order are observed. (b)-(d) Harmonic spectra generated in a 500-nm-thick Si sample on a 0.5-mm-thick sapphire substrate. (b) The solid black line represents the spectrum generated by a chirped, ~43 fs, synthesized pulse, where the odd and even harmonics up to 17$^{th}$ order are observed. The blue dotted line represents the spectrum generated by only the signal pulse, where only odd harmonics up to 13$^{th}$ order are observed. (c) The angle dependence of the harmonic spectra about the Si axis <001> relative to the laser polarization. The 4-fold symmetry confirms that the harmonic spectra originate from HHG in Si. The HHG yield is maximized when laser polarization is along <001>. (d) The near-continuous harmonic spectrum generated by the synthesized sub-cycle pulse without chirping. The harmonic comb structure is almost washed out due to the isolated emission from HHG.

In conclusion, we demonstrated a high-energy, CEP-stable, sub-cycle, mid-IR pulse synthesizer based on an OPA covering the bandwidth from 2.5 to 9.0 μm and drove HHG in thin Si samples to show the isolated sub-cycle strong-field interactions in solids. The synthesized pulse width was measured as ~12.4 fs, corresponding to 0.88 optical cycle at ~4.2 μm. The stable CEP was ensured with the passively CEP-stable pump and the WLG-seeded OPA scheme. 220 and 270 mrad rms CEP jitters were measured for the signal and idler pulses, respectively, over >6 minutes. A synthesized pulse with 33 μJ energy and ~1.9 GW peak power was obtained. HHG up to ~19$^{th}$ order was demonstrated in thin Si samples. The continuous harmonic spectrum confirmed that the generated sub-cycle mid-IR pulses allow for isolated harmonic emission. Further investigations in temporal domain along with CEP dependence are required for the detailed studies of strong-field electron dynamics in solids. We note that the energy scaling of this mid-IR source is relatively straightforward by adding OPA stages. Advanced pump laser technologies can potentially push the mid-IR sub-cycle source towards TW peak powers to drive strong-field interactions in gaseous media.

**Methods:**

**Mid-IR OPA construction.** An octave-spanning Ti:sapphire oscillator (650-1100 nm) generates the 2.1-µm signal seed through intrapulse DFG in a MgO:PPLN. The Ti:sapphire oscillator also seeds both *ps* 1047 nm Nd:YLF pump laser and *ps* 1030 nm cryo-cooled Yb:YAG pump laser. They provide ~1.5 mJ and ~45 mJ pump energy for the 2.1-µm OPCPA, respectively. The amplified 2.1-µm pulses from the 3-stage OPCPA are passively CEP stabilized with 3.5 mJ of maximum pulse energy and 26 fs pulse width. 0.8 mJ, 2.1-µm pulses are used as the pump of the mid-IR OPA. A 20 µJ portion of the 2.1-µm pump is split for WLG in a 6-mm-thick $BaF_2$ plate as the signal of the mid-IR OPA. The polarization is rotated by 90° using a $CaF_2$ half-wave plate to fulfill the type-I phase matching. A 1.1-mm-thick CSP from BAE Systems with $\theta=47°$ is chosen for the type-I parametric conversion for its large nonlinear coefficient, broad phase-matching bandwidth, and high damage threshold pumped by 2.1 µm pulses. As there is a lack of mid-IR broadband beam combiners and beam splitters, 300-µm thick Si plates are oriented at the Brewster angle of the 2.1 µm pump serving as the beam combiner and beam splitter. With a large refractive index (*i.e.*, $n\sim3.44$), Si has a large Brewster angle which is ~74° at 2.1 µm. This requires a ~74° incident angle of the signal and idler beams with orthogonal polarization to the 2.1-µm pump in the collinear OPA scheme, which gives ~70% reflection of the signal and idler pulses. The energy loss from the Si beam splitter is taken into account in Fig. 2(c).

**Spectral characterization.** The output spectra of the mid-IR OPA are recorded by a scanning-grating monochromator (Horiba) with a liquid-nitrogen cooled MCT detector. Long-pass filters with cutting wavelengths at 2400 nm, 3600 nm, 4500 nm, and 7300 nm are used to analyze the 2-octave-spanning spectrum.

**SHG FROG.** The temporal profile of the 2.1-µm pump and the amplified signal pulses is characterized using a SHG FROG appratus. A 100-µm thick *β*-barium borate (BBO) and a 140-µm thick $AgGaS_2$ is used for the SHG of the 2.1-µm pump and the amplified signal, respectively.

**XFROG.** A beam with 10 µJ energy is split from the 2.1-µm pump as the reference of the XFROG. The synthesized pulse and the reference pulse are focused on a 30-µm-thick type-I GaSe crystal. The dispersion of the crystal is as low as 2.5 $fs^2$ at 4.2 µm. The generated sum-frequency signal is coupled into a spectrometer with InGaAs detectors (NIRQuest512, Ocean Optics). The spectral resolution of 6.7 nm limits the overall resolution of the measured XFROG trace. An iris and a polarizer are used to block the leakage of the fundamental pulses and thus increase the signal to noise ratio. The delay scan is done at the synthesized pulse arm. For the XFROG measurement of an optimally synthesized pulse, the time delay between the signal and pump pulses is finely tuned using a piezo stage within a total delay of ~ 30 fs, such that the shortest duration is obtained while the amplified energy is maintained at maximum. The grid size of XFROG retrieval is 128.

**The cross-referencing *f-3f* nonlinear spectral interferometry.** We use the CEP-stable 2.1-μm pump beam as the CEP reference for the idler. By slightly rotating the CSP crystal and thereby inducing very small birefringence to the 2.1-μm pump, we can generate the weak, vertically-polarized component of the residual pump in Fig. 1 (~5 μJ of energy out of 0.8 mJ) while keeping the same performance of the mid-IR OPA. This polarization-rotated residual pump is also reflected by the Si beam splitter and collinear with the idler pulse which spans from 4.4 to 9.0 μm. Since the spectrum of the polarization-rotated residual pump is found to be narrower than the original pump spectrum, it is first focused to a 1-mm-thick ZGP crystal for the moderate spectral broadening through self-phase modulation. After that the collinear pulses, including the broadened residual pump and the idler, are focused to another 0.5-mm-thick ZGP to generate the TH of the idler in the wavelength of 1.7 to 2.4 μm, with the optimized phase-matching angle of the ZGP crystal. The spectrally overlapped TH of the idler and the broadened residual pump with a fixed time delay are coupled into a spectrometer with InGaAs detectors (NIRQuest512, Ocean Optics) for the cross-referencing SI measurement. The measured spectra of the broadened polarization-rotated residual pump and the TH of the idler are shown in Fig. S5(b). The single-shot interference measurements are done with an integration time of 1 ms, as presented in Fig S5(c).

**HHG setup.** The laser beam is focused using an *f*=25.4 mm off-axis parabolic mirror with gold coating and then collimated using an *f*=50.8 mm off-axis parabolic mirror with UV-enhanced Al coating which can reflect the high harmonics up to the UV range. The mid-IR and harmonic beams are focused into an UV-visible monochromator (SP-300i, Acton Research Corporation) with an ICCD camera (PI-MAX, Princeton Instruments). Low-order harmonics in the NIR region are not measured because the cutoff region is of main interest. A 200-nm-thick free-standing Si (<100>) sample and a 500-nm-thick Si (<100>) sample on a 0.5-mm-thick sapphire substrate are used as the medium of HHG. Each solid sample is mounted on a rotation stage and a translation stage for optimization of the HHG signal. The laser beam is focused into the sample with the normal angle of incidence and the vertical polarization. The crystal orientation relative to the laser polarization is rotated using the rotation stage while the position of the sample relative to the beam focus along the propagation direction is adjusted using the linear stage. While the focused spot sizes of signal and idler beams are estimated as ~13 μm and ~26 μm in Gaussian waist, respectively, with $M^2$ value of ~1.7, the sample position is optimized for the HHG experiment as explained in Section HHG in solids.

19. Blaga C. I., Xu J., DiChiara A. D., Sistrunk E., Zhang K., Agostini P., Miller T. A., DiMauro L. F. & Lin C. D., "Imaging ultrafast molecular dynamics with laser-induced electron diffraction," Nature **483**, 194-197 (2012).
20. Kling M. F., Siedschlag Ch., Verhoef A. J., Khan J. I., Schultze M., Uphues Th., Ni Y., Uiberacker M., Drescher M., Krausz F. & Vrakking M. J. J., "Control of electron localization in molecular dissociation," Science **312**, 246-248 (2006).
21. Brida D., Marangoni M., Manzoni C., Silvestri S. D. & Gerullo G., "Two-optical-cycle pulses in the mid-infrared from an optical parametric amplifier," Opt. Lett. **33**, 2901-2903 (2008).
22. Mayer B. W., Phillips C. R., Gallmann L., Fejer M. M. & Keller U., "Sub-four-cycle laser pulses directly from a high-repetition-rate optical parametric chirped-pulse amplifier at 3.4 µm," Opt. Lett. **38**, 4265-4268 (2013).
23. Mitrofanov A. V., Voronin A. A., Sidorov-Biryukov D. A., Mitryukovsky S. I., Fedotov A. B., Serebryannikov E. E., Meshchankin D. V., Shumakova V., Ališauskas S., Pugžlys A., Panchenko V. Ya., Baltuška A., and Zheltikov A. M., "Subterawatt few-cycle mid-infrared pulses from a single filament," Optica **3**, 299-302 (2016).
24. Pupeza I., Sánchez D., Zhang J., Lilienfein N., Seidel M., Karpowicz N., Paasch-Colberg T., Znakovskaya I., Pescher M., Schweinberger W., Pervak V., Fill E., Pronin O., Wei Z., Krausz F., Apolonski A. & Biegert J., "High-power sub-two-cycle mid-infrared pulses at 100 MHz repetition rate," Nature Photon. **9**, 721-724 (2015).
25. Krogen P., Suchowski H., Liang H. K., Kärtner F. X. & Moses J., "Mid-IR pulse shaping by adiabatic difference frequency conversion," NLO 2015 (OSA NLO conference) proceeding, NM3A.3 (2015) doi:10.1364/NLO.2015.NM3A.3.
26. Hemmer M., Baudisch M., Thai A., Couairon A. & Biegert J., "Self-compression to sub-3-cycle duration of mid-infrared optical pulses in dielectric", Opt. Express **21**, 28095-28102 (2013).
27. Liang H. K., Krogen P., Grynko R., Novak O., Chang C.-L., Stein G. J., Weerawarne D., Shim B., Kärtner F. X. & Hong K.-H., "Three-octave-spanning supercontinuum generation and sub-two cycle self-compression of mid-infrared filaments in dielectrics," Opt. Lett. **40**, 1069-1072 (2015).
28. Lanin A. A., Voronin A. A., Stepanov E. A., Fedotov A. B. & Zheltikov A. M., "Multioctave, 3-18 µm sub-two-cycle supercontinua from self-compressing, self-focusing soliton transients in a solid," Opt. Lett. **40**, 974-977 (2015).
29. Shumakova V., Malevich P., Ališauskas S., Voronin A., Zheltikov A. M., Faccio D., Kartashov D., Baltuška A. & Pugžlys A., "Multi-millijoule few-cycle mid-infrared pulses through nonlinear self-compression in bulk," Nature Comm. **7**, 12877-12892 (2016).
30. Nomura Y., Shirai H., Ishii K., Tsurumachi N., Voronin A. A., Zheltikov A. M. & Fuji T., "Phase-stable sub-cycle mid-infrared conical emission from filamentation in gases," Opt. Express **20**, 24741-24747 (2012).
31. Stepanov E. A., Lanin A. A., Voronin A. A., Fedotov A. B. & Zheltikov A. M., "Solid-state source of subcycle pulses in the midinfrared," Phy. Rev. Lett. **117**, 043901 (2016).
32. Fuji T., Nomura Y., and Shirai H., "Generation and Characterization of Phase-Stable Sub-Single-Cycle Pulses at 3000 cm$^{-1}$," IEEE J. Sel. Top. Quantum Electron. **21**, 8700612 (2015).
33. Krauss G., Lohss S., Hanke T., Sell A., Eggert S., Huber R. & Leitenstorfer A., "Synthesis of a single cycle of light with compact erbium-doped fibre technology," Nature Photon. **4**, 33–36 (2010).
34. Huang S.-W., Cirmi G., Moses J., Hong K.-H., Bhardwaj S., Birge J. R., Chen L.-J., Li E., Eggleton B. J., Cerullo G. & Kärtner F. X., "High-energy pulse synthesis with sub-cycle waveform control for strong-field physics," Nature Photon. **5**, 475-479 (2011).
35. Cerullo G. & Silvestri S. D., "Ultrafast optical parametric amplifiers," Rev. Sci. Instrum. **74**, 1-18 (2003).
36. Baumgarten C., Pedicone M., Bravo H., Wang H., Yin L., Menoni C. S., Rocca J. J., and Reagan B. A., "1 J, 0.5 kHz repetition rate picosecond laser," Opt. Lett. **41**, 3339 (2016).
37. Schunemann P. G. & Zawilski K. T., "Large aperture CSP for high-energy mid-infrared generation," OSA Congress on High-Brightness Sources and Light-Driven Interactions (EUV, HILAS, MICS) (Long Beach, CA, Mar. 20-22, 2016) MS2C.1.
38. Sanchez D., Hemmer M., Baudisch M., Cousin S. L., Zawilski K., Schunemann P., Chalus O., Simon-Boisson C. & Biegert J., "7 µm, ultrafast, sub-millijoule-level mid-infrared optical parametric chirped pulse amplifier pumped at 2 µm," Optica **3**, 147-150 (2016).
39. Malevich P., Kanai T., Hoogland H., Holzwarth R., Baltuška A. & Pugžlys A., "Broadband mid-infrared pulses from potassium titanly arsenate/znic germanium phosphate optical parametric amplfier pumped by Tm, Ho-fiber-seeded Ho:YAG chirped-pulse amplifier," Opt. Lett. **41**, 930-933 (2016).

**Acknowledgement:**

We thank Dr. T. Lang for letting us use the 2+1 dimensional nonlinear pulse propagation analyzer. We also thank Dr. O. D. Mücke for the theoretical support on solid HHG and Dr. P. D. Keathley for technical discussions on laser stability measurements. This work was supported by AFOSR (FA9550-12-1-0499, and FA9550-14-1-0255), the Center for Free-Electron Laser Science, DESY, Hamburg, Germany, and the excellence cluster "The Hamburg Centre for Ultrafast Imaging—Structure, Dynamics and Control of Matter at the Atomic Scale" of the Deutsche Forschungsgemeinschaft. H. K. Liang acknowledges the financial support from Singapore Institute of Manufacturing Technology (U14-P- 044AU and U14-P- 045AU) and A*STAR. P. K. acknowledges support from a NDSEG Graduate Fellowship. Z. W., H. P. and L. F. D. acknowledge support from the Air Force Office of Scientific Research under MURI, Award No. FA9550-16-1-0013. Z. W. acknowledges the support from the Presidential Fellowship of The Ohio State University.